\documentclass[conference]{IEEEtran}
\usepackage{subfigure}
\usepackage{amsmath,amssymb,amsfonts}
\usepackage{algorithm}
\usepackage[noend]{algpseudocode}

\usepackage{array}
\usepackage{textcomp}
\usepackage{stfloats}
\usepackage{url}
\usepackage{verbatim}
\usepackage{graphicx}
\usepackage{cite}
\usepackage[dvipsnames]{xcolor}

\usepackage{hyperref}
\hypersetup{
    colorlinks=true,
    linkcolor=blue,
    filecolor=black,      
    urlcolor=blue,
    citecolor=blue,
}
\usepackage[top=0.7in, bottom=0.7in, left=0.7in, right=0.7in]{geometry}

\begin{document}

\title{Digital Twin-Assisted Adaptive Multi-Agent DRL for Intelligent Spectrum and Resource Management in Open-RAN UAV-Enabled 6G Networks}
\author{
    \IEEEauthorblockN{Marwan Dhuheir, Thang X. Vu, and Symeon Chatzinotas,
    }
    \IEEEauthorblockA{The Interdisciplinary Centre for Security, Reliability and Trust (SnT), University of
Luxembourg, Luxembourg.
   }
}
\maketitle

\begin{abstract}
The evolution toward 6G wireless networks envisions a seamlessly intelligent, Open-RAN-enabled architecture where unmanned aerial vehicles (UAVs) play a pivotal role in extending coverage, enhancing resilience, and ensuring reliable connectivity for ground users’ deployment. However, efficiently managing spectrum and resources in such highly dynamic UAV-assisted environments remains a major challenge due to nonlinear system interactions, mobility-induced topology variations, and stringent latency and energy constraints. To address these challenges, we propose a digital twin (DT)-assisted adaptive deep reinforcement learning (DRL) framework that enables intelligent spectrum sharing and resource allocation across distributed ground users. The complex optimization problem is decomposed into UAV trajectory optimization using particle swarm optimization (PSO) and dynamic spectrum-power-association management via multi-agent DRL (MADRL). This hybrid DT-driven approach empowers intelligent, context-aware decision-making and adaptive coordination among UAVs. Extensive simulations demonstrate significant gains in spectral efficiency, data rates, and energy utilization, showcasing a transformative path toward self-evolving, autonomous 6G UAV and ground users (GUs) connectivity.
\end{abstract}


\section{introduction}
The evolution toward beyond 5G (B5G) and 6G wireless networks marks a paradigm shift in connectivity, targeting ultra-high data rates, extreme reliability, low latency (xURLLC), massive device connectivity, and AI-native intelligence across all layers \cite{she2023guest,9020161}. Achieving these goals demands scalable, adaptive, and intelligent network architectures capable of real-time optimization under dynamic conditions. A key enabler is the open radio access network (Open-RAN), which decouples hardware and software components across distributed units (DUs), radio units (RUs), central units (CUs), and RAN intelligent controllers (RICs), promoting flexibility and interoperability \cite{10528242}.
Complementing Open-RAN, digital twin (DT) technology creates real-time virtual replicas of physical network elements, enabling predictive analytics and proactive optimization \cite{masaracchia2023digital}. Integrated within the Open-RAN hierarchy, the DT enhances coordination between the Non-RT RIC (in the O-Cloud) and the Near-RT RIC (at the edge), enabling centralized training and distributed inference for intelligent self-optimization and low-latency orchestration.

Meanwhile, UAVs have emerged as crucial components of 6G non-terrestrial networks (NTNs) due to their mobility, rapid deployment, and ability to extend coverage to remote or disaster-stricken areas \cite{10978758}. Acting as aerial RUs (ARUs), UAVs support ground users (GUs) deployments and emergency communication, while DT-assisted Open-RAN control allows real-time flight path adjustment, dynamic resource allocation, and adaptive spectrum management. However, large-scale UAV-GU integration introduces challenges in spectrum sharing, energy efficiency, and maintaining reliable, low-latency links \cite{11161898,nguyen2024emerging}.
Existing works often rely on traditional optimization or game-theoretic methods \cite{11062573,10122731,10835238}, which ignore considering real-time adaptability, latency constraints, and concurrent coordination required in 6G networks. The study in \cite{10628007} adopted the DT for resource allocation, nevertheless, critical constraints hve not been considered in the evaluations and the model is limited to terrestrial network. Another related study in \cite{cao2025uav} studied the DT and DRL optimization for target search technology. However, the considered model is simple to adopt for improving wireless communication systems. To overcome these limitations, learning-driven mechanisms such as multi-agent reinforcement learning (MARL) have gained attention for enabling decentralized and adaptive decision-making \cite{11007537,tran2024deep}. When integrated with DT-based centralized training, MARL enables a DT-assisted MARL framework that combines global optimization through the Non-RT RIC with real-time distributed control via the Near-RT RIC, ensuring scalability, fast convergence, and robustness to dynamic environments.

To this end, we propose an adaptive DT-assisted multi-agent deep reinforcement learning (MADRL) framework for intelligent spectrum sharing and resource allocation in Open-RAN UAV-assisted 6G networks. The problem of data rate maximization is decomposed into two subproblems: UAV position optimization using low-complexity particle swarm optimization (PSO) and resource optimization (association, power, bandwidth) using adaptive MADRL. Unlike conventional methods, the proposed hybrid framework leverages DT-based centralized training for global synchronization and decentralized execution for real-time adaptability. The main contributions of this paper are summarized as follows:
\begin{itemize}
\item We propose a UAV-assisted Open-RAN DT-based framework for intelligent spectrum sharing and resource management in 6G wireless networks, addressing practical issues such as latency, energy consumption, and dynamic adaptability.
\item We formulate the joint data rate maximization problem by optimizing UAV positions, RU transmission power, bandwidth allocation, and RU-GU associations under the constraints of energy consumption, latency, and DT synchronization.
\item Due to the nonconvexity of the problem, we introduce an adaptive and iterative hybrid solution combining PSO for trajectory optimization and MADRL for decentralized resource allocations.
\item Extensive simulations validate the proposed framework, demonstrating superior performance over state-of-the-art solutions in terms of spectral efficiency, data rate maximization, and energy utilization.
\end{itemize}

The rest of this article is organized as follows: Section \ref{info_system_model} presents the description of our system model. In Section \ref{problem_formulation}, we delineate the problem formulation. Section \ref{Performance_evaluation} explains the implementation results of the proposed approach. At the end, section \ref{conclusion} concludes and discusses future research directions.

\section{system model}
\label{info_system_model}
\begin{figure}[t]
    \centering
\includegraphics[width=0.4\textwidth]{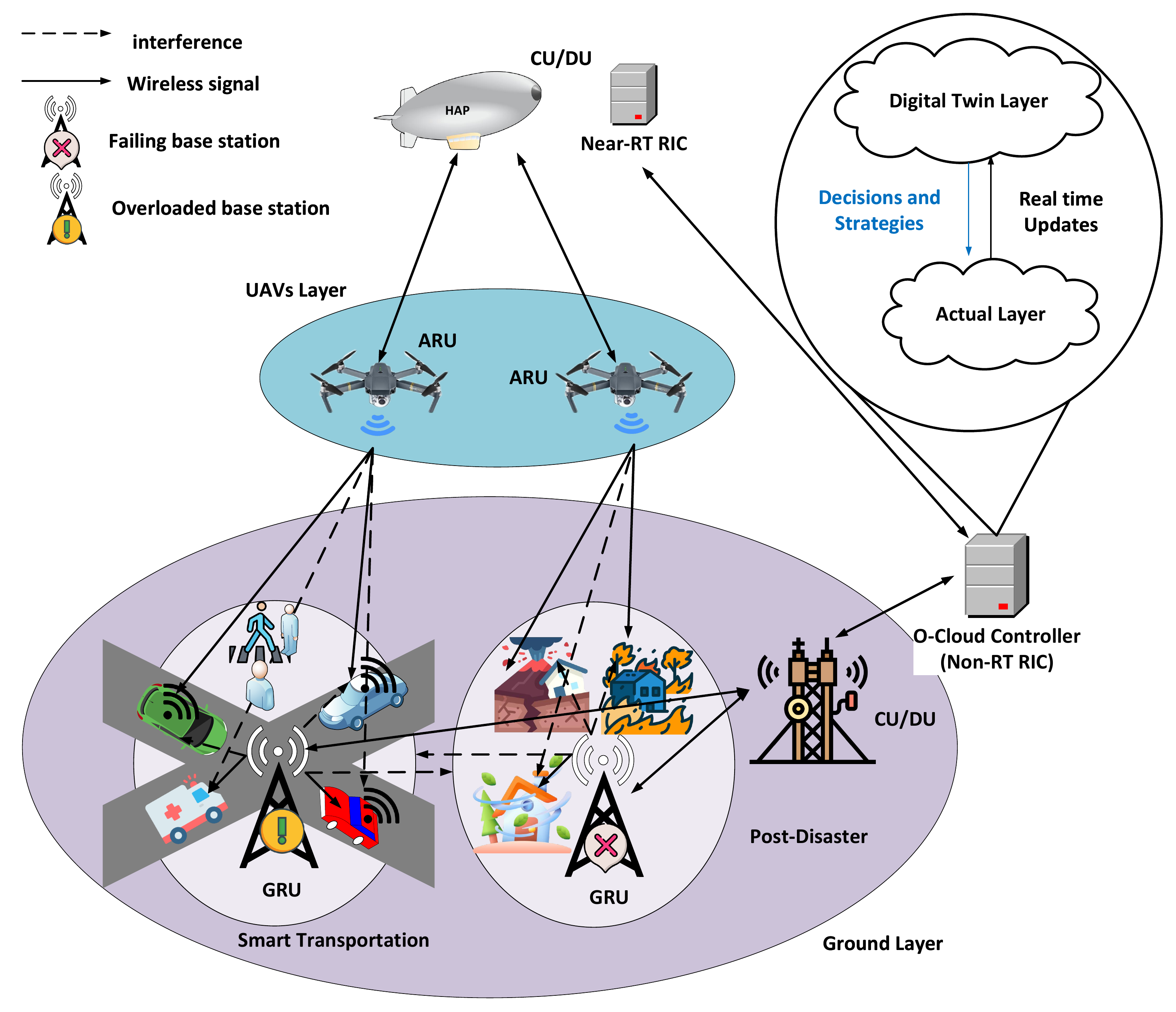}
    \caption{System model.}
    \label{System_Model}
\end{figure}
Fig.~\ref{System_Model} illustrates the proposed system model, which consists of a high-altitude platform (HAP), an Open-RAN CU/DU, multiple UAVs, ground radio units (GRUs), and distributed GUs operating within an Open-RAN-enabled 6G architecture. In this framework, some GRUs may become overloaded or experience service failures; therefore, UAVs are deployed as aerial radio units (ARUs) to assist GRUs in maintaining reliable connectivity for GUs such as post-disaster, special events, etc. Both RUs (UAVs and GRUs) provide wireless access and data services to the set of GUs $\mathcal{N} = \{1, 2, \ldots, N\}$ distributed across the coverage area.

Let $\mathcal{U} = \{1, 2, \ldots, U\}$ and $\mathcal{G} = \{1, 2, \ldots, G\}$ denote the sets of UAVs and GRUs, respectively, where $u \in \mathcal{U}$ and $g \in \mathcal{G}$. The network area is partitioned into clusters, each centered at the position of a GRU, such that every UAV supports one GRU cluster. The set of RU clusters is defined as $ru \in \mathcal{RU} = \{(GRU_1, UAV_1), (GRU_2, UAV_2), \ldots, (GRU_g, UAV_u)\}$, representing the pairwise association between GRUs and UAVs. Each UAV operates within its assigned cluster at a fixed altitude, following a trajectory that starts and ends at the corresponding cluster's center to ensure coordinated and energy-efficient operation. The positions of GRUs, UAVs, and GUs are denoted as $Q_g^t = (x_g^t, y_g^t, H_g)$, $Q_u^t = (x_u^t, y_u^t, H_u)$, and $Q_n = (x_n, y_n, 0)$, respectively.

During the service duration $T$, divided into $I$ discrete time slots $\mathcal{T} = \{1, 2, \ldots, I\}$ of length $\delta = T/I$, each UAV and GRU dynamically allocates its bandwidth and power resources to serve the associated GUs. 
We assume that UAVs share the available spectrum of GRUs to enhance spectral efficiency and overall throughput while ensuring interference coordination across aerial and terrestrial links. In the backhaul layer, GRUs connect to the Open-RAN CU/DU via terrestrial fiber or microwave links, whereas UAVs maintain aerial backhaul through the HAP. The HAP and CU/DU are further connected to the O-Cloud, which hosts the Non-RT RIC and the DT. The DT continuously receives real-time network data, including channel states, UAVs positions, and traffic loads through telemetry feedback from the RU, and the Near-RT RIC. Using the received data, the DT performs global training, policy optimization, and predictive analytics, generating updated control policies that are transmitted to the Near-RT RIC. These outputs enable real-time spectrum sharing, resource allocation, and adaptive UAV-GRU coordination at the network edge. Through this hierarchical Open-RAN-DT interaction, the system achieves continuous learning and low-latency, energy-efficient network management.

\subsection{Digital Twin Model}
The DT operates as a real-time virtual replica of the actual Open-RAN UAV-GU network, enabling continuous monitoring, prediction, and intelligent control through a tightly synchronized exchange between virtual and actual layers. Hosted in the O-Cloud, the DT collaborates with the non-RT RIC to perform centralized data aggregation, model training, and long-term optimization. The trained models, representing optimal control policies $\pi_u^*(a_u|s_u)$ for each agent $u \in \mathcal{U}\cup\mathcal{G}$ are periodically transmitted to the near-RT RIC, which deploys them for real-time inference and adaptation at the network edge. At each time step $t$, the DT receives state information from the actual layer, $D_{actu}^t=\{Q_g^t,Q_u^t,Q_n^t,\Gamma_{ru,n}^t\}$, to update its virtual counterpart $D_{DT}^t=f(D_{actu}^t)$. The reward output of the previous timestep $t-1$, computed from the actual system’s feedback (e.g., achieved throughput, latency, and energy consumption), is communicated back to the DT for refining the global model during centralized training. The DT updates its environment states continuously, while model synchronization between the non-RT and near-RT RICs occurs at slower, periodic timescales which typically every few seconds, allowing the near-RT RIC to react in milliseconds to real-time network variations using the most recently trained model. By following the hierarchical and asynchronous update structure, it ensures stable, scalable, and adaptive coordination across Open-RAN layers for efficient spectrum sharing and resource allocation.

\subsection{Energy Consumptions Model}
In UAV-assisted Open-RAN 6G networks, energy management plays a critical role in sustaining continuous operations, as UAVs are inherently constrained by limited onboard battery capacity. The total energy expenditure of a UAV is mainly determined by its propulsion mechanism and communication functions. However, the communication-related energy consumption is relatively negligible compared to propulsion energy and is thus omitted from the energy model. Accordingly, the UAV’s total energy consumption is dominated by propulsion energy, which depends on its flight dynamics, including velocity, altitude, and payload weight. To accurately represent this, the propulsion-power model for rotary-wing UAVs, as presented in \cite{9513250}, is adopted and expressed as follows:
\begin{equation}
\begin{aligned}
\varepsilon_{prop,u} = \underbrace{\eta_i \sqrt{\Bigg( \sqrt{\Big(1+\frac{v_u^4}{4v_0^4}}\Big)-\frac{v_u^2}{2v_0^2}\Bigg)} }_\textbf{Induced Power} + \underbrace{\eta_b \Big(1+\frac{3v_u^2}{v_{tip}^2}\Big) }_\textbf{Blade Power}\\ + \underbrace{\frac{1}{2}f_0\varphi r D_a v_u^3.}_\textbf{Parasite Power} \end{aligned} 
\label{prop_energy}
\end{equation}
In this model, $\eta_i$ and $\eta_b$ represent the induced and blade profile power coefficients, respectively; $v_{tip}$ denotes the rotor blade tip speed; $v_0$ is the rotor-induced velocity in hover conditions; $f_0$ is the fuselage drag ratio; $r$ denotes the rotor solidity; $\varphi$ represents the air density; and $D_a$ is the rotor disc area. The hovering power consumption can be derived from \eqref{prop_energy} by setting $v_u = 0$, as given by $\varepsilon_{hov,u} = \eta_i + \eta_b$.

Hence, the total energy consumption of UAV $u$ at time slot $t$ is defined as:
$\varepsilon_{u,tot}^t =
\varepsilon_{prop,u} \times t, \text{if } v_u > 0,
\varepsilon_{hov,u} \times t, \text{if } v_u = 0$.
Given their limited battery resources, UAVs must ensure sufficient energy reserves to complete their missions safely. The battery status of UAV $u$ at time slot $t$, denoted by $\Omega_u^t$, is updated as follows $\Omega_u^t = \Omega_u^{t-1} - \varepsilon_{u,tot}^t$,
where $\Omega_u^{t-1}$ represents the remaining battery level at the previous time slot. Let $\Omega_u^{0}$ denote the initial total battery capacity, such that $\Omega_u^{0} = \Omega_u^{init} + \Omega_u^{min}$, where $\Omega_u^{init}$ is the energy allocated for the operational mission and $\Omega_u^{min}$ is the minimum required energy for the UAV to safely return to its docking or charging station, which is the center of the cluster. Consequently, the UAV battery state is maintained within the feasible range $\Omega_u^t \in [\Omega_u^{min}, \Omega_u^{0}]$, ensuring energy-aware operation throughout the mission duration.

\subsection{End-to-End Service Latency Model}
The end-to-end latency in the proposed model comprises three components determining the overall service delay. The first is the access latency between RUs (UAVs/GRUs) and their associated GUs within each cluster, expressed as $\tau_{ru-n}^t=\frac{K_n^t}{R_{n,ru}^t}$, where $K_n^t$ is the data size requested by GU $n$ and $R_{n,ru}^t$ is the achievable data rate. The second component is the backhaul latency between the RUs and their aggregation nodes (HAP for UAVs or MBS for GRUs), given by $\tau_{ru-h,m}^t=\frac{K_u^t}{R_{ru,b}^t}$. The third is the cloud coordination and DT latency between HAP/MBS and the O-Cloud, expressed as $\tau_{h,m-oc}^t=\frac{K_{oc}^t}{R_{oc}^t}+\tau_{DT}^{dev}$, where $\tau_{DT}^{dev}$ represents the deviation due to DT synchronization delays. Thus, the total end-to-end latency is $\tau_{tot,n}^t=\tau_{ru-n}^t+\tau_{ru-h,m}^t+\tau_{h,m-oc}^t$.
To ensure reliable and timely transmission, the latency must satisfy $\tau_{tot,n}^t\leq\tau_{\max},\quad \forall n\in\mathcal{N},t\in\mathcal{T}$,
where $\tau_{\max}$ is the maximum tolerable delay threshold ensuring URLLC-grade reliability. During distributed execution, UAVs and GRUs utilize the DT-trained policies for rapid inference with minimal on-site delay, while predictive synchronization and proactive updates compensate for $\tau_{DT}^{dev}$, guaranteeing low-latency and reliable service delivery.

\subsection{Wireless Communication Model}

We consider that the RUs (UAVs and GRUs) communicate with GUs through OFDM over $M$ subchannels. All RUs and the HAP share the same spectrum resources and operate within the same set of subchannels $\mathcal{M} = \{1, 2, \ldots, M\}$ to ensure efficient spectrum utilization and coordinated interference management. The movement of UAVs is constrained by their maximum permissible velocity $V_{\max}$ (m/s) and their consumed energy, and thus their positions must satisfy the following mobility constraint at each time slot $t \in \mathcal{T}$ as $(x_u^{t+1}-x_u^t)^2 + (y_u^{t+1}-y_u^t)^2 \leq V_{\max}^2, \quad \forall u \in \mathcal{U}$.

To prevent collisions and maintain safe inter-UAV spacing, a minimum distance constraint is imposed as $(x_u^t - x_v^t)^2 + (y_u^t - y_v^t)^2 \geq d_{\min}^2, \quad \forall u,v \in \mathcal{U},u \neq v$.
The communication links between RUs (UAVs/GRUs) and GUs are modeled according to their propagation nature. 
UAV-GU air-to-ground channels follow a Rician fading model with a dominant LoS component, whereas GRU-GU terrestrial links follow a Rayleigh fading model with rich multipath scattering. 
The instantaneous channel power gain is given by
$g_{n,ru}^t = |h_{n,ru}^t|^2 L_{n,ru}^t, \quad 
L_{n,ru}^t = \frac{d_0}{(d_{n,ru}^t)^\alpha}, \;
d_{n,ru}^t = \sqrt{(x_n - x_{ru}^t)^2 + (y_n - y_{ru}^t)^2 + H_{ru}^2}$.
For UAV-GU links, 
$h_{n,ru}^t = \sqrt{\frac{K}{K+1}}\psi_{\text{LoS}} + \sqrt{\frac{1}{K+1}}\psi_{\text{NLoS}}$, 
while for GRU-GU links, $h_{n,g}^t \!\sim\! \mathcal{CN}(0,1)$. 
The SINR and data rate are given by
$\Gamma_{n,ru}^t = 
\frac{P_{n,ru}^t a_{n,ru}^t g_{n,ru}^t}
{\sum_{{B_g}\in\mathcal{G}} a_{n,B_g}^t B_{n,B_g}^t g_{n,B_g}^t + N_0},$
where $a_{n,ru}^t \in \{0,1\}$ is the binary association variable indicating whether GU $n$ is served by RU $ru$ (UAV or GRU) at time $t$, 
$P_{n,ru}^t$ is the transmission power, and $N_0$ is the Gaussian noise power. 
The term $\sum_{{B_g} \in \mathcal{G}} a_{n,B_g}^t B_{n,B_g}^t g_{n,B_g}^t$ represents the aggregate interference from neighboring to the GU. 
Accordingly, the achievable data rate between GU $n$ and RU $ru$ is formulated as:
\begin{equation}
R_{n,ru}^t = B_{n,ru}^t \log_2(1 + \Gamma_{n,ru}^t),
\label{data_rate}
\end{equation}
where $B_{n,ru}^t$ denotes the bandwidth allocated to GU $n$ by RU $ru$ at time $t$.

Considering that both UAVs and GRUs operate as RUs under Open-RAN coordination, their spectrum resources are shared adaptively based on instantaneous channel and traffic conditions. The service amount $S_{n,ru}^t$ represents the volume of data successfully transmitted from RU $ru$ to GU $n$ during time slot $t$, which depends on the achievable rate and slot duration $\delta^t$. It can be expressed as
\begin{equation}
S_{n,ru}^t = \delta^t R_{n,ru}^t, \quad \forall n \in \mathcal{N}, ru \in \mathcal{RU}, t \in \mathcal{T}.
\label{service_amount}
\end{equation}

In this model, the Open-RAN control architecture, through the Near-RT RIC, dynamically allocates transmission power, bandwidth, and subchannel resources among UAVs and GRUs, while the DT-assisted centralized layer continuously refines the channel and mobility models to enhance link reliability, spectrum efficiency, and latency-aware service delivery across the 6G network.

\section{Problem Formulation}
\label{problem_formulation}

The optimization problem is formulated to maximize the overall service provision to distributed GUs in an Open-RAN UAV-assisted 6G network. The primary decision variables include the UAVs’ positions, RUs-GUs associations, transmission power levels, and spectrum resource allocation indicators. Accordingly, the optimization framework can be expressed as
\begin{normalsize}
\begin{flalign}
    &\max_{\mathcal{A},Q, \mathcal{B},\mathcal{P}} \sum_{t \in \mathcal{T}} \sum_{ru \in \mathcal{RU}} \sum_{n \in \mathcal{N}} S_{n,ru}^t 
    \label{objective_fun}\\
    & \mathrm{subject~to:}\nonumber\\
    &C_1:\Omega_u^t \geq \Omega_u^{min}, \forall u \in \mathcal{U}, \forall t \in \mathcal{T}
    \tag{\ref{objective_fun}a}
         \label{C1}\\
    & C2: \tau_{tot,n}^t \leq \tau_{\max}, \quad \forall n \in \mathcal{N}, t \in \mathcal{T},
         \label{C2}
         \tag{\ref{objective_fun}b}\\
    & C3: \Gamma_{n,ru}^t \geq \Gamma_{n,ru}^{min}, \forall ru \in \mathcal{RU}, \forall n \in \mathcal{N}, \forall t \in \mathcal{T},
    \label{C3}
         \tag{\ref{objective_fun}c}\\
    & C4: 0 \leq B_{n,ru}^t \leq B_{max}, \forall ru \in \mathcal{RU}, \forall n \in \bar{N},
    \label{C4}
    \tag{\ref{objective_fun}d}\\
    & C5: 0 \leq P_{n,ru}^t \leq P_{max}, \forall ru \in \mathcal{RU}, \forall n \in \bar{N},
     \label{C5}
     \tag{\ref{objective_fun}e}\\
     & C6:  Q_u^0 = Q_u^L, \forall u \in \mathcal{U} 
     \label{C6}
     \tag{\ref{objective_fun}f}\\
    &  C7: 0 \leq x_u^t,y_u^t \leq x_{max},y_{max}, \forall u \in \mathcal{U}, \forall t \in \mathcal{T},
    \label{C7}
    \tag{\ref{objective_fun}g}\\
    &  C8: (x_u^{t+1}-x_u^t)^2 + (y_u^{t+1}-y_u^t)^2 \leq V_{max}^2, \forall u \in \mathcal{U},
    \label{C8}
    \tag{\ref{objective_fun}h}\\
    & C9: (x_u^t-x_v^t)^2 + (y_u^t-y_v^t)^2 \geq d_{min}^2, \forall u,v \in \mathcal{U}, u \neq v, 
    \label{C9}
    \tag{\ref{objective_fun}i}\\
    & C10: a_{n,ru}^t \in \{0,1\}, \forall ru \in \mathcal{RU}, \forall n \in \mathcal{N}, \forall t \in \mathcal{T},
    \label{C10}
    \tag{\ref{objective_fun}j}\\
    &  C11: \sum_{ru=1}^{RU} a_{n,ru}^t \leq 1, \forall ru \in \mathcal{RU}, \forall n \in \mathcal{N}, \forall t \in \mathcal{T},
    \label{C11}
    \tag{\ref{objective_fun}k}
\end{flalign}
\end{normalsize}

In this formulation, $\mathcal{A}={a_{n,ru}^t}$ represents the RU-GU association variables, $Q={x_u^t,y_u^t}$ denotes the UAV positions, $\mathcal{B}={B_{n,ru}^t}$ defines the allocated bandwidth resources, and $\mathcal{P}={P_{n,ru}^t}$ denotes the transmission power levels. The objective in Eq.~\eqref{objective_fun} aims to maximize the total service provision $S_{n,ru}$ to distributed GUs by jointly optimizing UAV trajectories, RU-GU associations, and power-spectrum allocation under Open-RAN coordination.

Constraint $C1$ ensures sufficient energy availability for each UAV throughout its mission. Constraint $C2$ enforces the end-to-end latency requirement for reliable and ultra-low-latency service provisioning. Constraint $C3$ guarantees that each communication link satisfies the minimum SINR threshold for reliable data transmission. Constraints $C4$-$C5$ define feasible bandwidth allocation, transmission power and ensure that total allocated resources remain within available limits.  Constraint $C6$ enforces closed-loop UAV trajectories such that the start and end positions coincide. Constraint $C7$ restricts UAV movement to the defined geographical boundaries, while $C8$ and $C9$ govern UAV mobility and collision avoidance. Finally, $C10$ enforces binary RU-GU association, and $C11$ ensures that each GU is served by at most one RU per time slot.

The optimization problem ~\eqref{objective_fun} is formulated as a mixed-integer nonlinear programming (MINLP) problem, characterized by high nonconvexity arising from binary association variables, UAV mobility coupling, and nonlinear channel dynamics, making traditional convex optimization approaches unsuitable for real-time operation. To overcome these challenges, we decompose the problem into two interrelated subproblems: UAV trajectory optimization, addressed via a lightweight PSO algorithm to enhance coverage and reduce interference, and resource management, solved through the adaptive MADRL for joint association, bandwidth, and power allocation, which will be described in the next section.

\section{The Proposed PSO and MADRL-based solutions}
\label{propsed_solution}
In the proposed hybrid optimization framework, UAV positioning and resource management are jointly optimized through an iterative PSO-MADRL procedure. At the beginning of each episode, the PSO algorithm determines feasible UAV trajectories by iteratively updating each particle’s position and velocity to maximize the objective in ~\eqref{objective_fun}, subject to UAV mobility and coverage constraints. The optimized UAV positions are then provided to the adaptive MADRL module, which refines RU-GU associations, bandwidth allocation, and transmission power based on the updated environment. As shown in Algorithm \ref{alg:DT_PSO_MADRL}, the two algorithms operate sequentially within each iteration, where PSO fixes UAV positions for the current service interval, while MADRL adapts resource policies dynamically according to observed states and rewards. Feasibility is ensured by enforcing trajectory, energy, and latency constraints at every iteration, and convergence is achieved as the joint optimization stabilizes when both UAV positions and MADRL policies yield minimal reward variance across consecutive episodes.
\begin{algorithm}[ht]
\caption{DT-assisted PSO-MADRL Framework}
\label{alg:DT_PSO_MADRL}
\begin{algorithmic}[1]
\small
\State \textbf{Initialize:} UAV/GRU positions $Q^0$, twin critics $Q_{\theta_1}, Q_{\theta_2}$, actor $\pi_\phi$, target networks $\bar{\theta}, \bar{\phi}$, replay buffer $\mathcal{D}$.
\For{each episode $e=1$ to $E$}
    \State \textbf{PSO phase:} optimize UAV positions $Q_u$ to maximize $S_{n,ru}^t$ under constraints (\ref{C1})-(\ref{C11}).
    \For{each time step $t \in \mathcal{T}$}
        \State Observe state $s_{ru}^t=\{Q_{ru}^t,\Omega_u^t,R_{n,ru}^t,\Gamma_{n,ru}^t,\tau_{\text{tot},n}^t\}$.
        \State Select action $a_{ru}^t=\pi_\phi(s_{ru}^t)+\mathcal{N}(0,\sigma_\theta^2)$ for $\{a_{n,ru}^t,B_{n,ru}^t,P_{n,ru}^t\}$.
        \State Execute action, update environment, and compute reward $r_{ru}^t=\omega_1R_{n,ru}^t-\omega_2\tau_{\text{tot},n}^t-\omega_3\varepsilon_{u,\text{tot}}^t$.
        \State Store $(s^t,a^t,r^t,s^{t+1})$ in $\mathcal{D}$.
        \If{update step}
            \State Compute target: $a'=\pi_{\bar{\phi}}(s')+\text{clip}(\epsilon,-c,c)$, $\epsilon\!\sim\!\mathcal{N}(0,\sigma^2)$.
            \State $y=r+\gamma\min(Q_{\bar{\theta}_1},Q_{\bar{\theta}_2})$; update critics $\theta_1,\theta_2$ and actor $\phi$.
            \State Soft update targets: $\bar{\theta}\!\leftarrow\!\tau\theta+(1-\tau)\bar{\theta}$, $\bar{\phi}\!\leftarrow\!\tau\phi+(1-\tau)\bar{\phi}$.
        \EndIf
    \EndFor
    \State \textbf{DT update:} synchronize trained parameters between non-RT (DT) and near-RT RICs for next episode.
\EndFor
\end{algorithmic}
\end{algorithm}
The resource optimization phase employs an enhanced deep deterministic policy gradient (DDPG) algorithm with architectural enhancements to improve learning stability, scalability, and robustness in dynamic environments. More specifically, twin critic networks $Q_{\theta_1}(s,a)$ and $Q_{\theta_2}(s,a)$ are trained concurrently, using the minimum of their predictions, $Q_{\text{target}}=\min(Q_{\theta_1},Q_{\theta_2})$, to counteract overestimation bias. Policy smoothness is further maintained through target policy smoothing, where Gaussian noise $\epsilon \sim \mathcal{N}(0,\sigma^2)$ is added to the target actor’s output, $a'=\pi_{\bar{\phi}}(s')+\text{clip}(\epsilon,-c,c)$, reducing sensitivity to sharp Q-value variations. Additionally, parameter noise adaptation is integrated into the actor network for state-dependent exploration, dynamically adjusted via the Kullback-Leibler (KL) divergence between perturbed and nominal policies to balance exploration and exploitation.


In the proposed hybrid DT-assisted PSO-MADRL framework, feasibility is maintained by enforcing all system constraints during both UAV trajectory optimization and resource allocation management. At the beginning of each episode, the PSO algorithm determines feasible UAV positions that satisfy mobility, collision avoidance, energy, and latency constraints based on the distribution of active GUs. The obtained positions remain fixed throughout the episode, during which the adaptive MADRL module-based on an enhanced DDPG architecture optimizes spectrum sharing, transmission power, and RU-GU associations while ensuring feasibility through projection and adaptive action refinement. Each RU agent (UAV or GRU) interacts locally with its environment, while the centralized critic in the DT layer, implemented through the non-RT RIC, aggregates global state-action information for coordinated training. The near-RT RIC executes decentralized actor policies for real-time decision-making. The iterative and independent coordination between PSO trajectory optimization and MADRL policy refinement enables stable convergence, scalability, and robust adaptability, ensuring reliable, low-latency, and energy-efficient operation across the considered Open-RAN UAV-assisted framework.

In the proposed MADRL framework, the resource management process is modeled as a Markov Decision Process (MDP), where each RU agent (UAV or GRU) interacts with the environment to learn optimal control policies. The MDP elements are described as follows:
\subsubsection{\textbf{states}} The state space of each agent $ru$ at time slot $t$ is defined as
$s_{ru}^t = \{Q_{ru}^t,Q_n^t, \Omega_u^t,\Gamma_{n,ru}^t, \tau_{\text{tot},n}^t\}$,
which includes the RUs positions, remaining energy, SINR, and latency.
\subsubsection{\textbf{actions}}  The action space $a_{ru}^t = \{a_{n,ru}^t, B_{n,ru}^t, P_{n,ru}^t\}$ consists of RU-GU association, bandwidth allocation, and transmission power. 
\subsubsection{\textbf{rewards}} The reward function is designed to guide cooperative optimization among agents and is formulated as
$r_{ru}^t = \sum_{n \in \mathcal{N}} R_{n,ru}^t$
\section{simulation results and analysis}
\label{Performance_evaluation}
The performance of the proposed DT-assisted PSO-MADRL framework is evaluated against four benchmark algorithms, MADDPG, MAPPO, MA Actor-Critic, and a Greedy solution to validate its efficiency in optimizing UAV positioning, spectrum sharing, and resource allocation. Simulations are conducted over a $1000 \text{m} \times 1000 \text{m}$ area containing 100 GUs served by three clusters (3 GRUs and 3 UAVs), whose spatial distribution follows a power-law pattern to emulate realistic, nonuniform device densities. The experiments are implemented in TensorFlow 1.13.1 and Python 3.7.16 on a Windows system with an Intel Xeon i7 (2.2 GHz, 4-core) and 16 GB RAM. Key parameters include UAV velocity $v_u=30$ m/s, noise power $\sigma^2=-95$ dBm, packet size $K_n=10$ MB, horizon $L=180$ s, discount factor $\gamma=0.99$, and carrier frequency $f_c=5$ MHz. The maximum latency, bandwidth, and power are set to $\tau_{\max}=100$ ms, $B_{\max}=100$ MHz, and $P_{\max}=35$ dBm, respectively, ensuring a realistic and dynamic evaluation of the model.
\subsection{Clusters and UAVs Movements}
\begin{figure}[t!]
    \centering
    \includegraphics[scale=0.25]{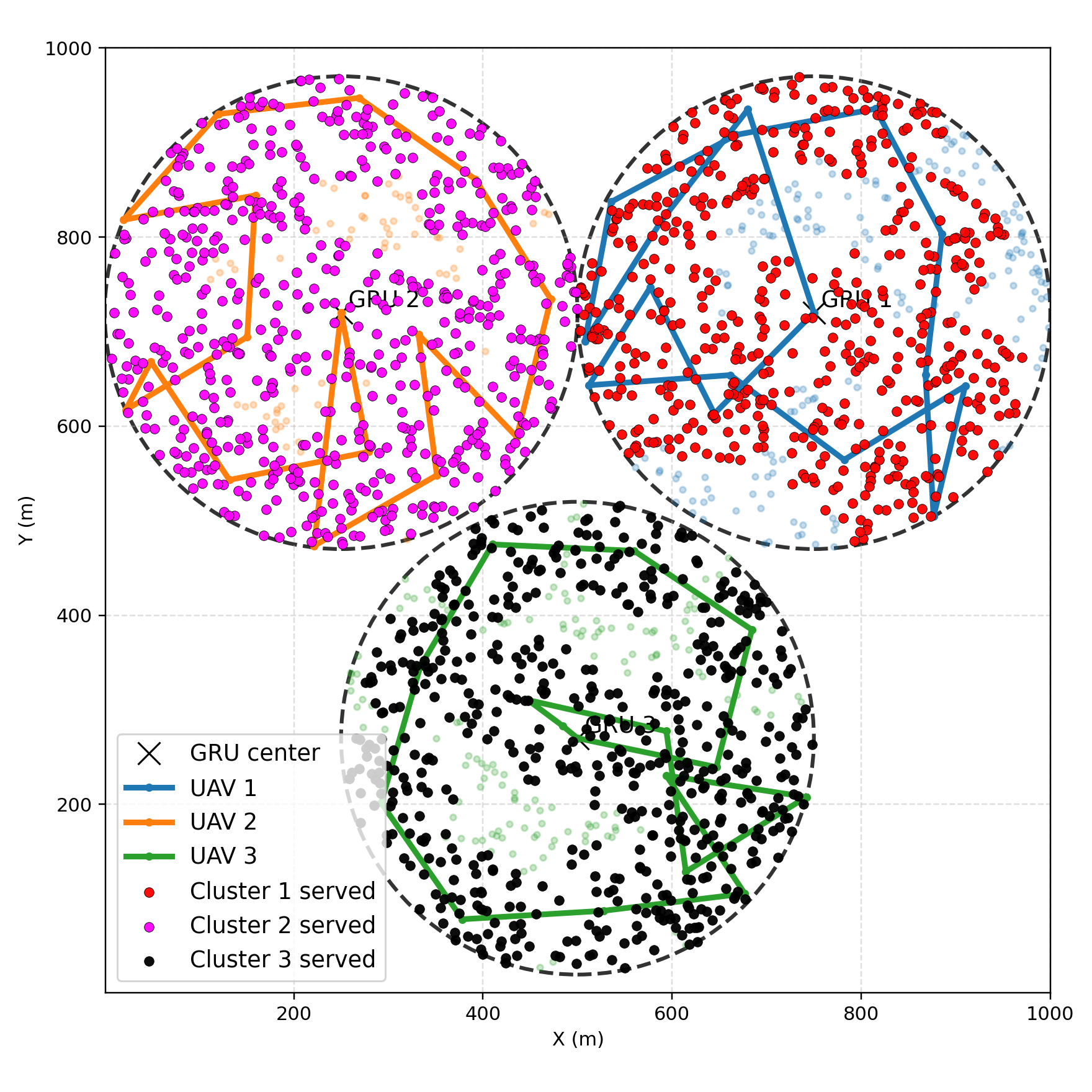}
    \caption{UAVs Paths within their serving cluster.}
    \label{UAVs_paths}
\end{figure}
Fig.~\ref{UAVs_paths} shows the optimized UAV trajectories and RU-GU associations, where each cluster is centered around a GRU and a UAV dynamically moves to serve distributed GUs. Using the PSO algorithm over 10 episodes, UAV movements are optimized to maximize data rates, enhance coverage, and reduce interference. The results demonstrate that UAVs and GRUs collaboratively achieve near-complete GU coverage within all clusters, ensuring efficient and reliable service delivery.
\subsection{Convergence Analysis}
Fig.~\ref{convergence} illustrates the convergence behavior of the proposed MADRL-based framework compared to benchmark algorithms. The proposed method achieves significantly faster convergence and higher average rewards, stabilizing after nearly 200 episodes, while other algorithms such as MA Actor-Critic, MADDPG, and MAPPO exhibit slower learning and lower reward saturation. This demonstrates the superior stability, learning efficiency, and adaptability of the proposed framework under dynamic Open-RAN UAV network conditions.
\subsection{Data Rate Analysis}
Figs.~\ref{cluster_data_rate} and \ref{data_rates} illustrate the average data rate performance of the proposed framework compared with benchmark algorithms in terms of convergence and changing the number of clusters. As shown in Fig.~\ref{cluster_data_rate}, the proposed DT-assisted PSO-MADRL approach consistently achieves higher average data rates across varying cluster numbers, highlighting its scalability and efficient spectrum utilization in dense UAV-GU deployments. Similarly, Fig.~\ref{data_rates} demonstrates the data rate convergence over training episodes, where the proposed model rapidly stabilizes at a higher throughput level compared to MA Actor-Critic, MADDPG, MAPPO, and the Greedy solution. These results confirm the proposed framework’s capability to jointly optimize UAV positioning, power control, and spectrum allocation for enhanced network performance.

\subsection{Latency Analysis}
\begin{figure*}[ht!]
\centering
	\mbox{
	    \hspace{-7mm} 
        \subfigure[\label{convergence}]{\includegraphics[scale=0.30]{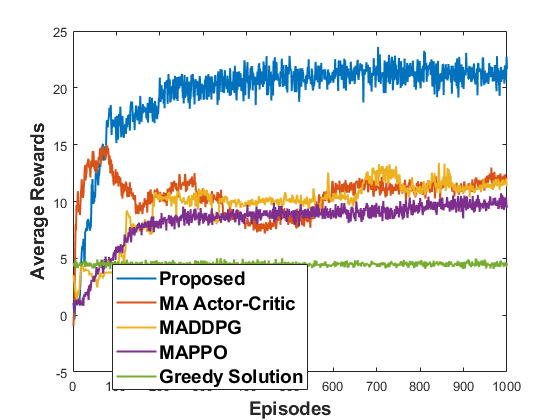}
	   }
	     \subfigure[\label{data_rates}]{\includegraphics[scale=0.30]{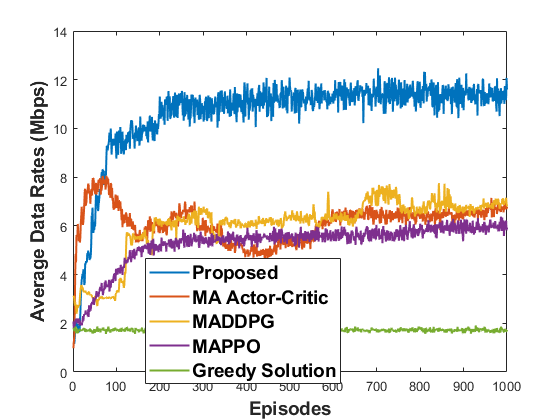}}
} 
	\mbox{
		\hspace{-7mm}
		\subfigure[\label{cluster_data_rate}]{\includegraphics[scale=0.17]{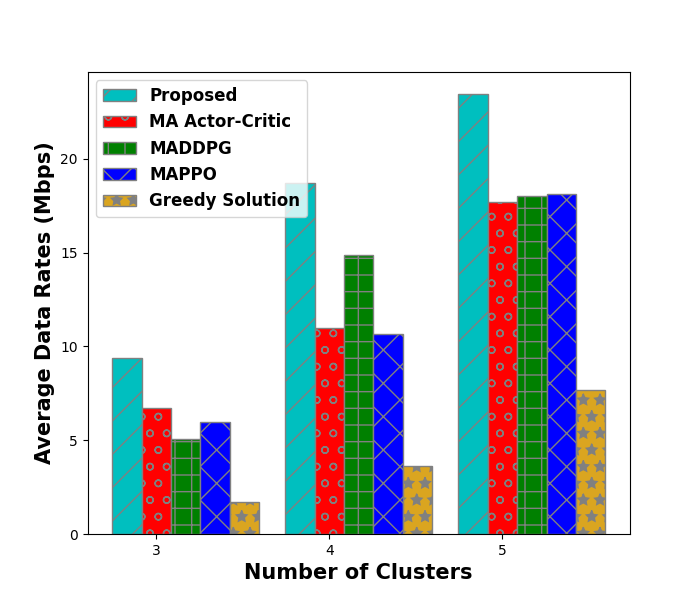}}
		\subfigure[\label{latency}]{\includegraphics[scale=0.30]{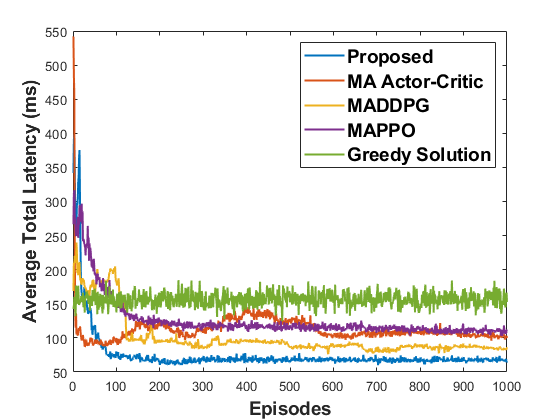}}
}
	\caption{Proposed solution comparison with benchmark algorithms in terms of convergence, data rate, and latency.}
	\label{fig:graph4}
\end{figure*}

Fig.~\ref{latency} presents the convergence of the average total latency over training episodes for the proposed framework and the benchmark algorithms. The proposed DT-assisted PSO-MADRL-based solution achieves the lowest and most stable latency, converging rapidly to around 60 ms, outperforming MA Actor-Critic, MADDPG, MAPPO, and the Greedy solution. This superior performance demonstrates the framework’s capability to ensure URLLC through efficient selection of UAVs positions, RUs-GUs associations, spectrum sharing, adaptive resource allocation, and DT-guided policy optimization.
\section{conclusion}
\label{conclusion}
This work tackled the challenge of intelligent spectrum sharing and resource allocation in Open-RAN UAV-enabled 6G networks, addressing the need for low-latency, energy-efficient, and high-throughput communication where GRUs are supported by UAVs to enhance GU connectivity. To overcome the limitations of traditional optimization methods, a DT-assisted adaptive MADRL framework was proposed, integrating PSO-based UAV trajectory optimization for stable, scalable, and robust learning. Extensive simulations verified that the proposed approach achieves higher data rates, faster convergence, and lower latency than existing benchmarks, demonstrating its effectiveness in enabling autonomous, adaptive, and intelligent 6G Open-RAN networks.

\section*{Acknowledgment}
This work was supported in whole, or in part, by the Luxembourg National Research Fund (FNR), ref. C22/IS/17220888/RUTINE and BRIDGES/2023/IS/18441334/Pre5GNR

\bibliographystyle{IEEEtran}
\bibliography{bibliography.bib}
\end{document}